\newcommand{\costht}{\cos{\theta}}

\newcommand{\dedx}{dE/dx}

\newcommand{\llb}{\Lambda\bar{\Lambda}}
\newcommand{\llp}{\llb\pi^0}
\newcommand{\lle}{\llb\eta}
\newcommand{\lm}{\Lambda}
\newcommand{\lmb}{\bar{\Lambda}}

\newcommand{\GG}{\gamma\gamma}

\newcommand{\splb}{\Sigma^+\pi^-\bar{\Lambda}}
\newcommand{\sbpl}{\bar{\Sigma}^-\pi^+\lm}
\newcommand{\SSB}{\Sigma^0\bar{\Sigma}^0}

\newcommand{\pppr}{\pi^+\pi^-p\bar{p}}

\newcommand{\psp}{\psi(2S)}
\newcommand{\jpsi}{J/\psi}
\newcommand{\ar}{\rightarrow}

\newcommand{\bfg}{\begin{figure}}
\newcommand{\efg}{\end{figure}}
\newcommand{\bitm}{\begin{itemize}}
\newcommand{\eitm}{\end{itemize}}
\newcommand{\bnum}{\begin{enumerate}}
\newcommand{\enum}{\end{enumerate}}
\newcommand{\btbl}{\begin{table}}
\newcommand{\etbl}{\end{table}}
\newcommand{\btbu}{\begin{tabular}}
\newcommand{\etbu}{\end{tabular}}
\newcommand{\bcl}{\begin{center}}
\newcommand{\ecl}{\end{center}}
\newcommand{\bbt}{\bibitem}
\newcommand{\beq}{\begin{equation}}
\newcommand{\eeq}{\end{equation}}
\newcommand{\beqr}{\begin{eqnarray}}
\newcommand{\eeqr}{\end{eqnarray}}

\documentclass[twocolumn,tightenlines,showpacs,amsmath,amssymb]{revtex4}
\usepackage{rotating,epsfig,graphicx}
\usepackage{flafter}
\include{def-com}
\normalsize
\begin{document}
\title{\boldmath \bf Measurements of $\jpsi$ and $\psp$ decays into $\llp$ and $\lle$ }
\author{
M.~Ablikim$^{1}$,              J.~Z.~Bai$^{1}$, Y.~Ban$^{12}$,
X.~Cai$^{1}$,                  H.~F.~Chen$^{17}$, H.~S.~Chen$^{1}$,
H.~X.~Chen$^{1}$, J.~C.~Chen$^{1}$, Jin~Chen$^{1}$,
Y.~B.~Chen$^{1}$
% ...
\\
\vspace{0.2cm}
(BES Collaboration)\\
\vspace{0.2cm}
{\it
$^{1}$ Institute of High Energy Physics, Beijing 100049, People's Republic of China\\
% ...
$^{12}$ Peking University, Beijing 100871, People's Republic of China\\
% ...
$^{17}$ University of Science and Technology of China, Hefei 230026, People's Republic of China\\
% ...
}
}
\date{\today}
\begin{abstract}
Using 58 million $\jpsi$ and 14 million $\psi(2S)$ events collected by
the BESII detector at the BEPC, branching fractions or upper limits
for the decays $\jpsi$ and $\psp\ar\llp$ and $\lle$ are measured. For
the isospin violating decays, the upper limits are determined to be
${\cal B}(\jpsi\ar\llp)<6.4\times 10^{-5}$ and ${\cal
B}(\psp\ar\llp)<4.9\times 10^{-5}$ at the 90\% confidence level. The
isospin conserving process $\jpsi\ar\lle$ is observed for the first
time, and its branching fraction is measured to be ${\cal
B}(\jpsi\ar\lle)=(2.62\pm 0.60\pm 0.44)\times 10^{-4}$, where the
first error is statistical and the second one is systematic. No $\lle$
signal is observed in $\psp$ decays, and ${\cal
B}(\psp\ar\lle)<1.2\times 10^{-4}$ is set at the 90\% confidence
level. Branching fractions of $\jpsi$ decays into $\splb$ and $\sbpl$
are also reported, and the sum of these branching fractions is
determined to be ${\cal B}(\jpsi\ar\Sigma^+\pi^-
\bar{\Lambda}+c.c.)=(1.52\pm 0.08\pm 0.16)\times 10^{-3}$.
\end{abstract}
\maketitle
\section{Introduction}
Several charmonium decay modes containing $\lm\lmb$ pairs have been
reported~\cite{np1,np2,np3,np4,np5,np6,np7,np8}. Among these decays,
the isospin violating process $\jpsi\ar\llp$ has been studied by
DM2~\cite{np3} and BESI~\cite{np4}, and its average branching fraction
is determined to be ${\cal B}(\jpsi\ar\llp)=(2.2\pm 0.6)\times
10^{-4}$~\cite{np12}. However, the isospin conserving process
$\jpsi\ar\lle$ has not been reported, and there are no measurements
for $\llp$ and $\lle$ decays of $\psp$.

In this paper, we study $\jpsi$ and $\psp\ar\llp$, $\lle$ using 58 M
$\jpsi$ events and 14 M $\psp$ events taken with the BESII detector at
the BEPC storage ring. We find that the $\jpsi\ar\llp$ branching
fraction is much smaller than those measured by DM2 and BESI. In
addition, we observe the isospin conserving process $\jpsi\ar\lle$ and
measure its branching fraction for the first time. Analyses of $\llp$
and $\lle$ in $\psp$ decays are also performed, but no obvious signals
are observed for these two channels.

\section{The BESII Detector and Monte Carlo simulation}
BESII is a conventional solenoidal magnet detector that is described
in detail in Ref.~\cite{np9}. A 12-layer vertex chamber (VTC)
surrounding the beam pipe provides trigger and track information. A
forty-layer main drift chamber (MDC), located radially outside the
VTC, provides trajectory and energy loss ($dE/dx$) information for
charged tracks over $85\%$ of the total solid angle.  The momentum
resolution is $\sigma _p/p = 0.0178 \sqrt{1+p^2}$ ($p$ in
$\hbox{GeV}/c$), and the $dE/dx$ resolution for hadron tracks is $\sim
8\%$. An array of 48 scintillation counters surrounding the MDC
measures the time-of-flight (TOF) of charged tracks with a resolution
of $\sim 200$ ps for hadrons.  Radially outside the TOF system is a 12
radiation length, lead gas-tube barrel shower counter (BSC). This measures
the energies of electrons and photons over $\sim 80\%$ of the total
solid angle with an energy resolution of $\sigma_E/E=21\%/\sqrt{E}$
($E$ in GeV). Outside of the solenoidal coil, which provides a
0.4~Tesla magnetic field over the tracking volume, is an iron flux
return that is instrumented with three double layers of counters that
identify muons of momentum greater than 0.5~GeV/$c$.

In this analysis, a GEANT3 based Monte Carlo (MC) simulation
program~\cite{np10} with detailed consideration of real detector
responses (such as dead electronic channels) is used. The consistency
between data and Monte Carlo has been carefully checked in many
high-purity physics channels, and the agreement is quite
reasonable~\cite{np11}.

\section{Event selection}
The decay channels investigated in this paper are $\jpsi\ar\llp$,
$\jpsi\ar\lle$, $\psp\ar\llp$, and $\psp\ar\lle$, where $\lm$ decays
to $\pi^- p$ and $\pi^0$ and $\eta$ to $\GG$. The final states in which
we are interested contain two photons and four charged tracks
($\pppr$). Candidate events are required to satisfy the following
common selection criteria:
\begin{enumerate}
\item Events must have four good charged tracks with net charge
zero. A good charged track is a track that is well fitted to a helix
in the MDC and has a polar angle, $\theta$, in the range
$|\costht|<0.8$.
\item The TOF and $\dedx$ measurements of the charged tracks are used
to calculate $\chi^{2}_{PID}$ values for the hypotheses that the
particle is a pion, kaon, or proton. Only the two proton tracks must
be identified with the requirement that $\chi^{2}_{PID}$ for
the proton hypothesis is less than those for the $\pi$ or $K$
hypotheses.
\item Isolated photons are those that have an energy deposit in the BSC
greater than 50 MeV and have the angle between the photon entering the BSC,
and the shower development direction in the BSC less than
$37^{\circ}$. In order to remove the fake photons produced by
$\bar{p}$ annihilation and those produced by hadronic interactions of
tracks with the shower counter, the angle between the photon and
antiproton is required to be larger than $25^{\circ}$ and those
between the photon and other charged tracks larger than $8^{\circ}$.
\item The selected events are subjected to four constraint (4C)
kinematic fits. When there are more than two candidate photons in an
event, all combinations are tried and the combination with the
smallest $\chi^{2}_{4C}$ is retained. The selection requirement on
$\chi^{2}_{4C}$ is optimized by maximizing $S/\sqrt{S+B}$, where
$S$ and $B$ are the expected numbers of signal and background events,
respectively.
\item To select $\lm$ and $\lmb$, the difference between the measured
$\pi p$ mass and the expected mass ($M(\lm)$) should be less than
10 MeV/$c^2$ (three times the $\lm$ mass resolution).
\end{enumerate}

\section{Event analysis}
\subsection{\boldmath $\jpsi\ar\llp$}
\subsubsection{Event Selection}
Only events with two good photons are selected, and 4C kinematic fits
under the $\GG \pppr$ hypothesis are performed. To select clean
$\llb$ events, we require the $\lm$ and $\lmb$ secondary vertices to be
reconstructed successfully, and the decay lengths of $\lm$ and $\lmb$
in the $x-y$ plane must be larger than 0.05 m.

\bfg
\centerline{\psfig{file=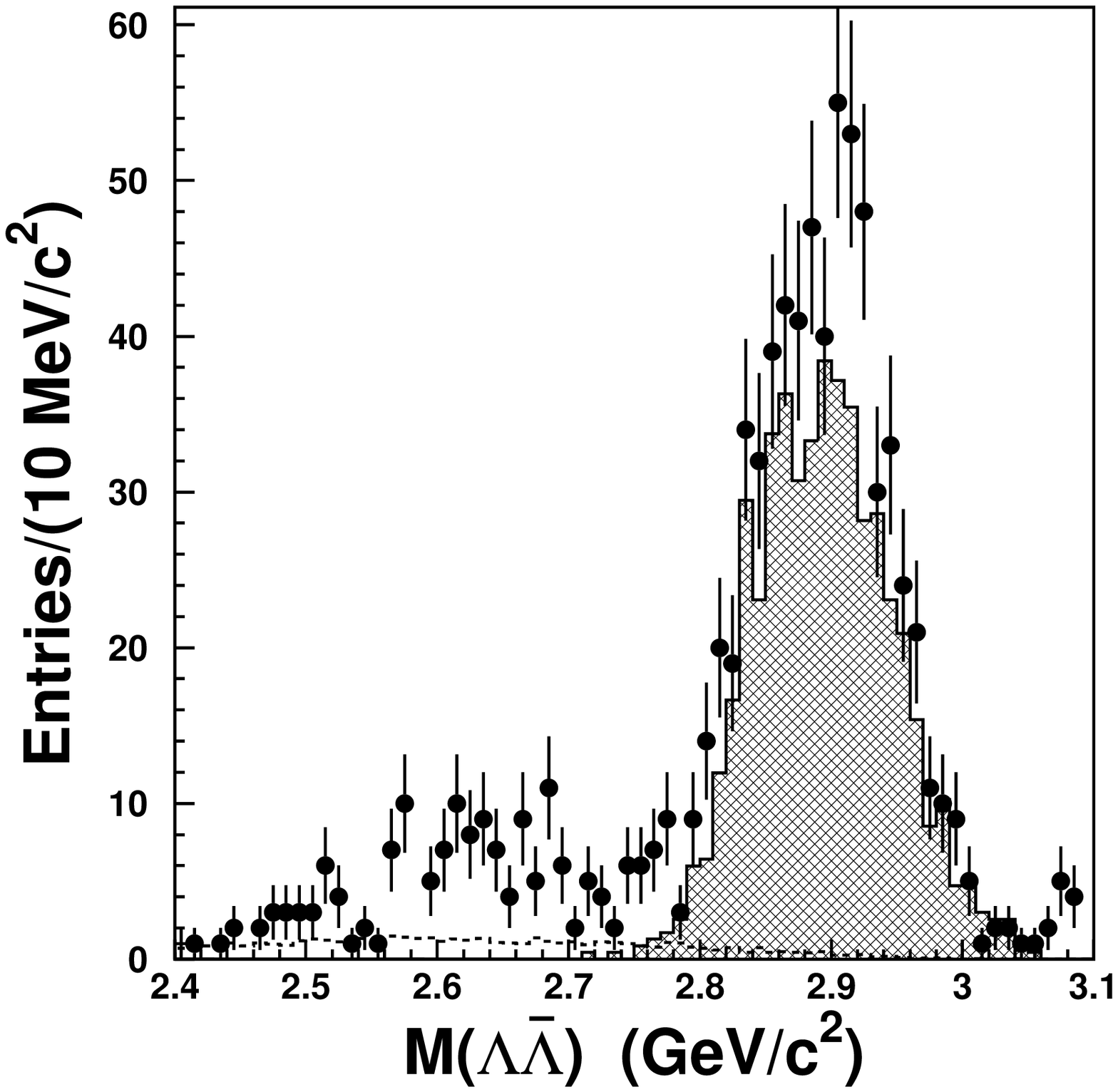,width=8cm,height=6cm} }
\caption{Distribution of $M(\llb)$ for $\jpsi\ar\llb\GG$ candidates.
Dots with error bars are data, the shaded histogram is background from
MC simulated $\jpsi\ar\SSB$, normalized according to the branching
fraction in the PDG, and the dashed histogram is the MC simulated
$\jpsi\ar\llp$ signal,  normalized according to the branching
fraction in the PDG.}
\label{nnfig1}
\efg

Figure ~\ref{nnfig1} shows the $\llb$ invariant mass ($M(\llb)$)
distribution after
the above selection. The large peak near 2.9 GeV/$c^2$ is background
from $\jpsi\ar\SSB$, in agreement with the expectation from the MC
simulation, normalized to its branching fraction~\cite{np12}, shown as
the shaded histogram in Fig.~\ref{nnfig1}. To reject such background,
$M(\llb)$ is required to be less than 2.8 GeV/$c^2$.  With this
selection, Fig.~\ref{nnfig3} shows the $\chi^{2}_{4C}$ distribution
for data and Monte Carlo simulation. To suppress potential
backgrounds, $\chi^{2}_{4C}<$ 10 is required.

\bfg
\centerline{\psfig{file=newfig1/nfig2.epsi,width=8cm,height=6cm}}
\caption{Distribution of $\chi^{2}_{4C}$ for $\jpsi\ar\llb\GG$
  candidate events (solid histogram) and Monte Carlo simulated $\jpsi\ar\llp$ events
  (dashed histogram). Here, $M(\llb)$ is required to be less than 2.8
  GeV/$c^2$.}
\label{nnfig3}
\efg

\bfg
\centerline{\psfig{file=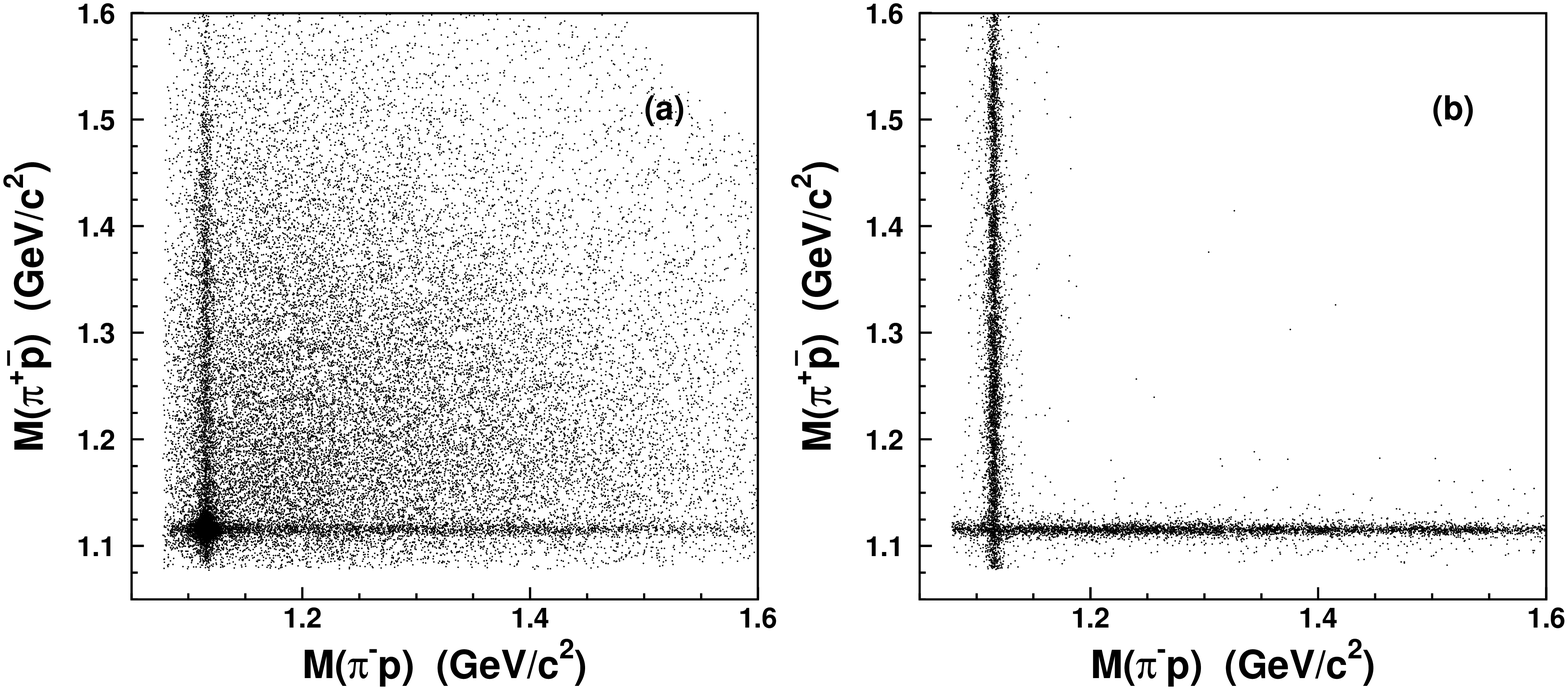,width=8cm,height=4cm}}
\caption{Scatter plot of  $M(\pi^+ \bar{p})$ versus
$M(\pi^- p)$ invariant mass for (a) $\jpsi\ar\splb
~(+c.c.)$ candidate events and (b) MC simulation.}
\label{nnfig14}
\efg
\subsubsection{Background Analysis}
To explore other possible backgrounds, we generate MC events for the
following channels: $\jpsi\ar\gamma\llb$, $\SSB$,
$\Sigma(1385)^0\bar{\Sigma}(1385)^0$, $\Xi^0\bar{\Xi}^0$,
$\Xi(1530)^0\bar{\Xi}^0$, $\Sigma^0\pi^0\lmb+c.c.$, and
$\Sigma^+\pi^-\lmb+c.c.$. Only the last two channels give significant
contributions to the $\pi^0$ signal. In particular, the decay mode
$\jpsi\ar\Sigma^0\pi^0\lmb+c.c.$, which contains $\llp$ with an
additional photon in the final state, could contribute
to the observed number of $\llp$ candidates. Because direct
measurements of $\jpsi\ar\Sigma^0\pi^0\lmb+c.c.$ are difficult, we measure the
branching fractions of their isospin partners and estimate their
branching fractions by assuming isospin symmetry. To estimate the
contamination from $\jpsi\ar\Sigma^0\pi^0\lmb+c.c.$, a high precision
measurement of $\jpsi\ar\Sigma^+\pi^-\lmb+c.c.$ is very important.

\bfg
\centerline{\psfig{file=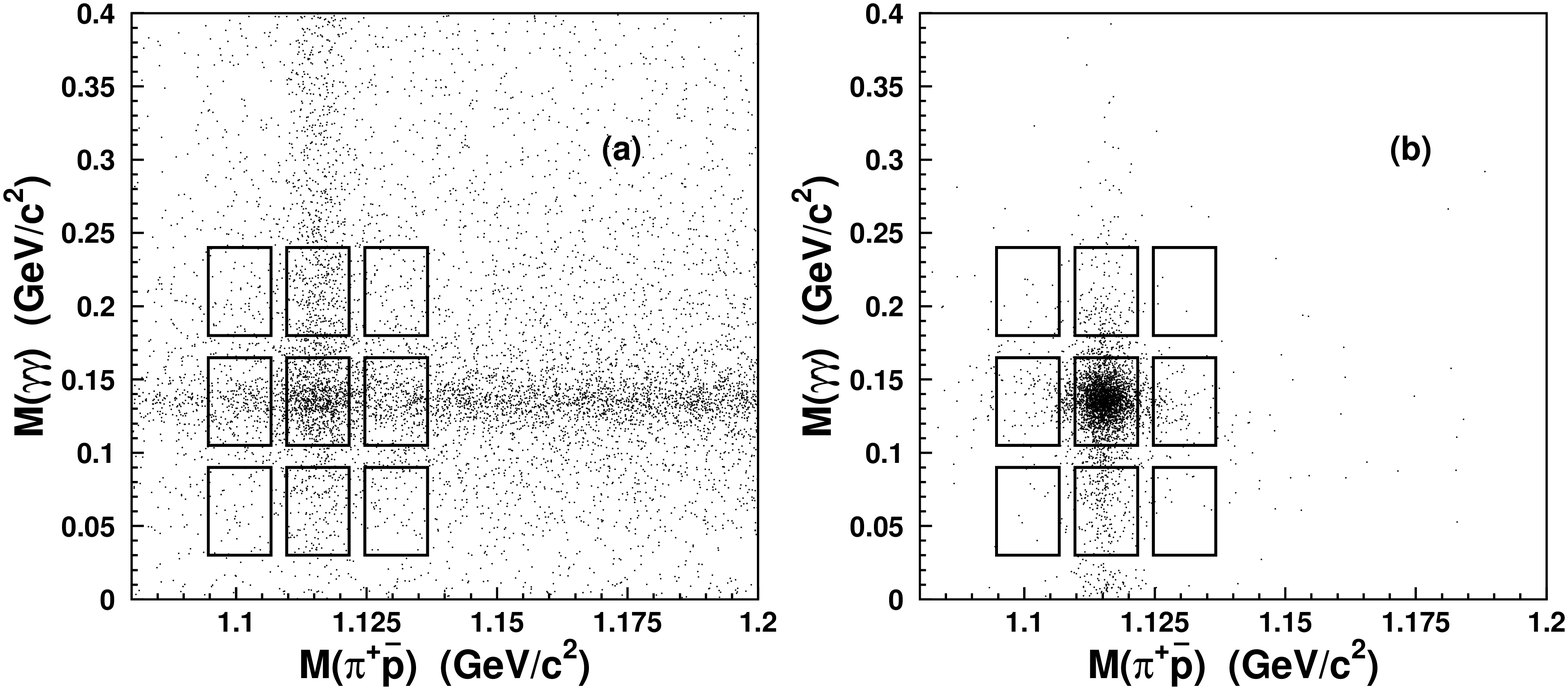,width=8cm,height=4cm}}
\caption{Scatter plot of $M(\GG)$ versus $M(\pi^+ \bar{p})$ for (a)
  $\jpsi\ar\splb$ candidate events and (b) MC simulation of
  $\jpsi\ar\splb$, both satisfying $\chi^{2}_{4C}<$15. The central box in
  the figure is the signal region defined by $|M(\GG)-M(\pi^0)|<0.03$
  GeV/$c^2$ and $|M(\pi^+ \bar{p})-M(\lmb)|<0.006$ GeV/$c^2$. The
  $\pi^0$ sideband is defined by $|M(\GG)-0.06|<0.03$ MeV/$c^2$ and
  $|M(\GG)-0.21|<0.03$ GeV/$c^2$ (The two boxes located above and
  below the signal region), and the $\lmb$ sideband region by
  $|M(\pi^+ \bar{p})-1.101)|<0.006$ GeV/$c^2$ and $|M(\pi^+
  \bar{p})-1.131)|<0.006$ GeV/$c^2$ (The two boxes on the left and
  right of the signal region). The four boxes at the corners are used
  to estimate the phase space contribution.}
\label{nnfig15}
\efg

\bfg
\centerline{\hbox{\psfig{file=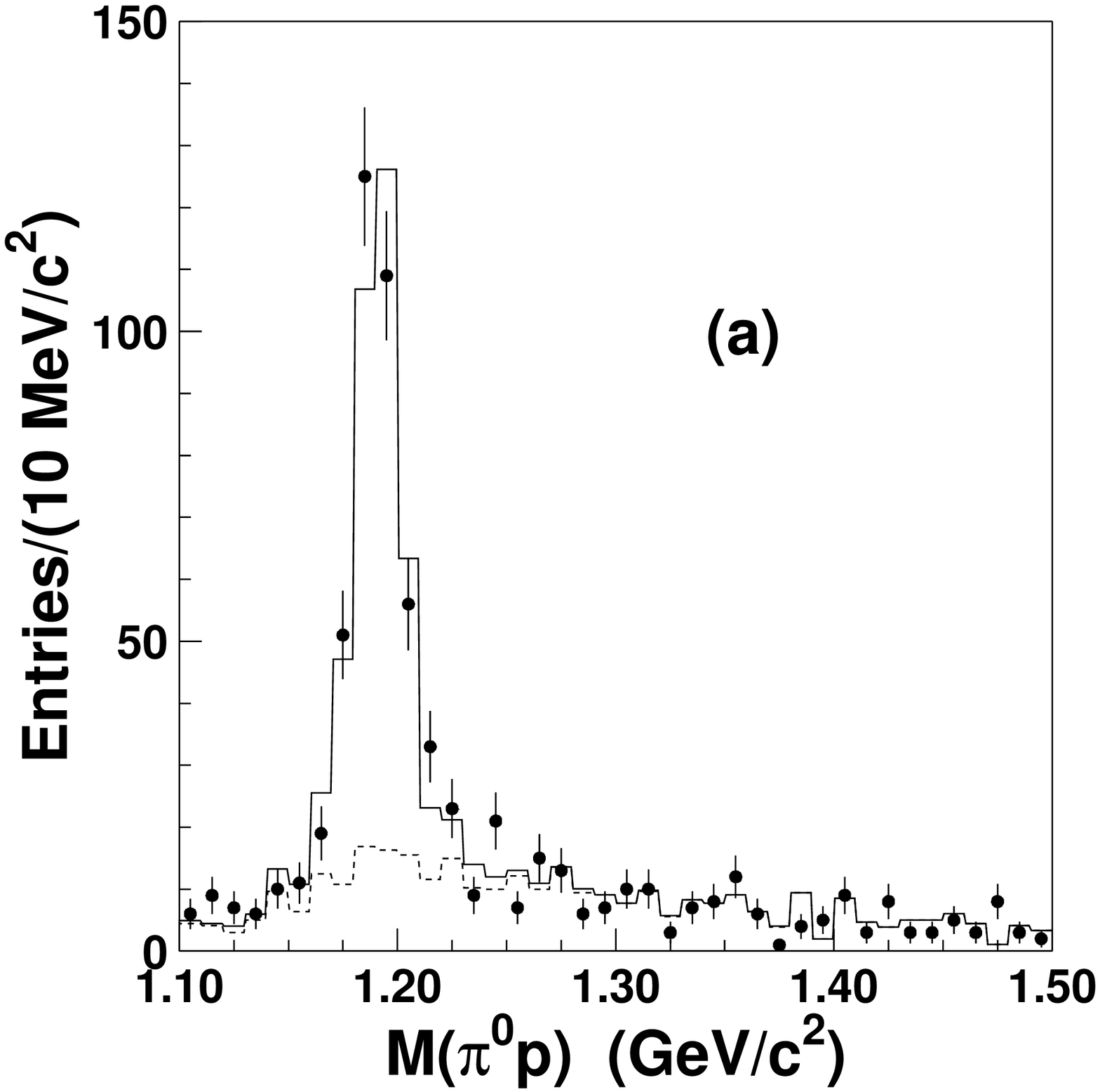,width=4cm,height=4cm}
\psfig{file=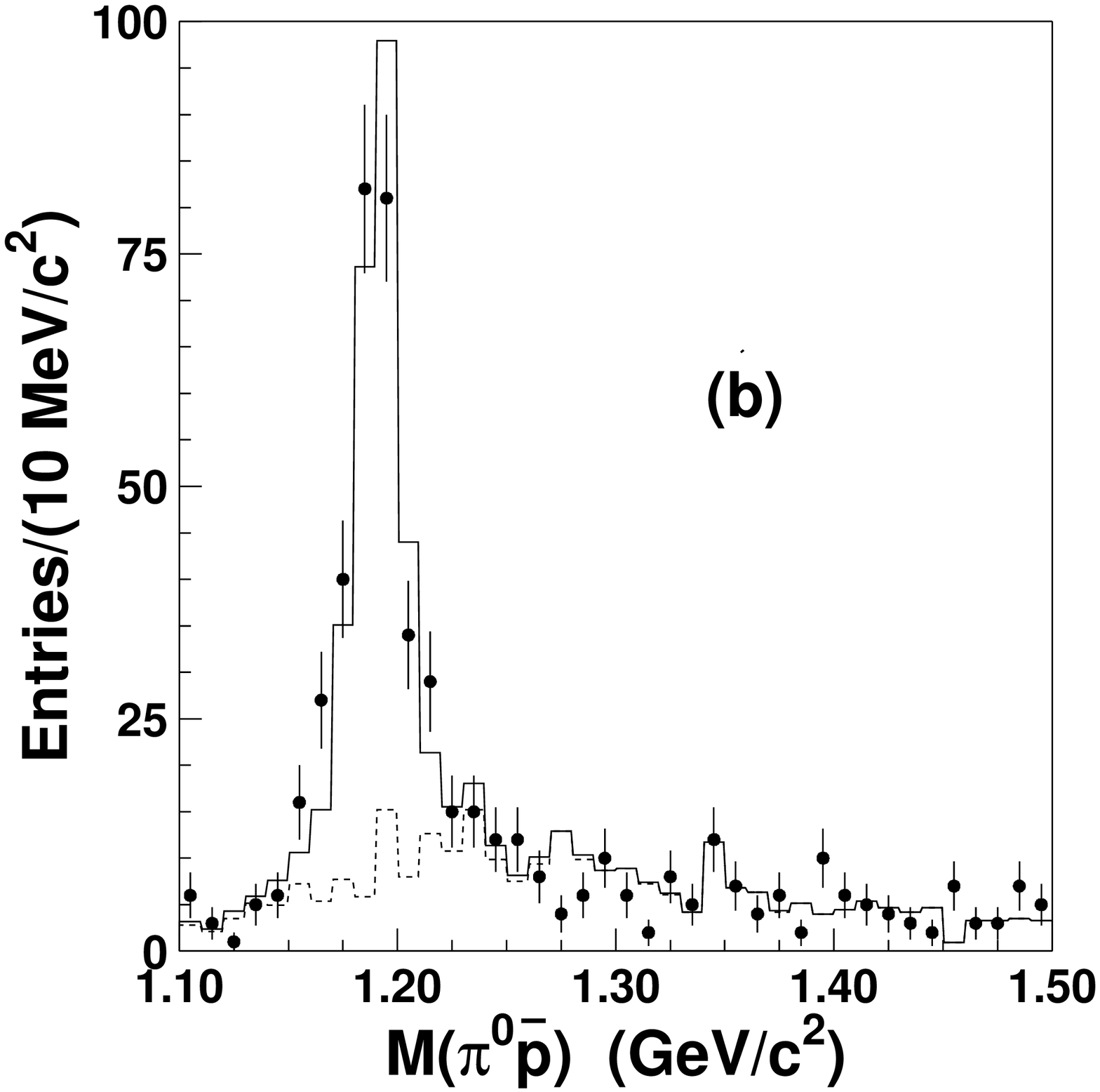,width=4cm,height=4cm}}}
\caption{(a) $M(\pi^0p)$ of $\jpsi\ar\pi^0p\pi^-\lmb$ candidate
  events from the signal region of Fig.~\ref{nnfig15} and (b)
  $M(\pi^0\bar{p})$ of $\jpsi\ar\pi^0\bar{p}\pi^+\lm$ candidate
  events. Dots with error bars are data, the solid histograms are the
  best fits described in the text, and the dashed histograms are
  backgrounds estimated from  $\lm$ and $\pi^0$ sidebands.}
\label{nnfig17}
\efg
\subsubsection{Measurement of $\jpsi\ar\Sigma^+\pi^-\lmb+c.c.$}
The $\jpsi\ar\splb+c.c.$ events, where $\Sigma^+ \ar
\pi^0 p$ and $\lmb\ar\pi^+ \bar{p}$, have the same final states as the
signal channel $\llp$. Candidate events are required to satisfy
$\chi^{2}_{4C}<$15, in addition to the common selection criteria in
Section III, except for the $\lm$ and $\lmb$ mass
requirements. Figure~\ref{nnfig14} is a scatter plot of $M(\pi^+
\bar{p})$ versus $M(\pi^- p)$ invariant mass for data and MC
simulation. The two bands are the $\jpsi\ar\splb+c.c.$ events. In
order to select $\jpsi\ar\splb$, $M(\pi^- p)>$ 1.15 GeV/$c^2$ is
required. Figure~\ref{nnfig15} shows the scatter plot of $M(\GG)$ versus
$M(\pi^+ \bar{p})$. The intersection region (central box) of the
$\pi^0$ and $\lmb$ bands corresponds to the $\jpsi\ar\splb$
signal. The dots with error bars in Fig.~\ref{nnfig17} (a) show the
distribution of $M(\pi^0 p)$ invariant mass of the events in the
central box ($|M(\GG)-M(\pi^0)|<30$ MeV/$c^2$ and $|M(\pi^+
\bar{p})-M(\lm)|<6$ MeV/$c^2$), and a clear $\Sigma^+$ signal is
observed. The dashed histogram is the background coming from sidebands
of $\pi^0$ and $\lmb$. To obtain the number of $\Sigma^+$ events, we
fit the $\Sigma^+$ signal with a histogram of the signal shape from MC
simulation plus the background shape determined from the $\pi^0$ and
$\lmb$ sidebands. $335\pm22$ $\Sigma^+$ events are obtained from the
fit. We do a similar analysis to measure $\jpsi\ar\sbpl$. The
signal for $\bar{\Sigma}^-$ and the fitting result are shown in
Fig.~\ref{nnfig17} (b). The fit yields $254 \pm 19$ events.

The efficiencies for $\jpsi\ar\splb$ and $\jpsi\ar\sbpl$ are determined
to be 2.3\% and 1.8\% using $2\times 10^{5}$ MC simulated signal
events, respectively. The branching fractions are calculated to be
${\cal B}(\jpsi\ar\splb)=(7.70\pm 0.51)\times 10^{-4}$ and ${\cal
B}(\jpsi\ar\sbpl)=(7.47\pm 0.56)\times 10^{-4}$, where the errors are
statistical. The total branching fraction of the two conjugate modes is ${\cal B}(\jpsi\ar\splb+c.c.)=(15.17\pm 0.76)\times 10^{-4}$.

\bfg
\centerline{\psfig{file=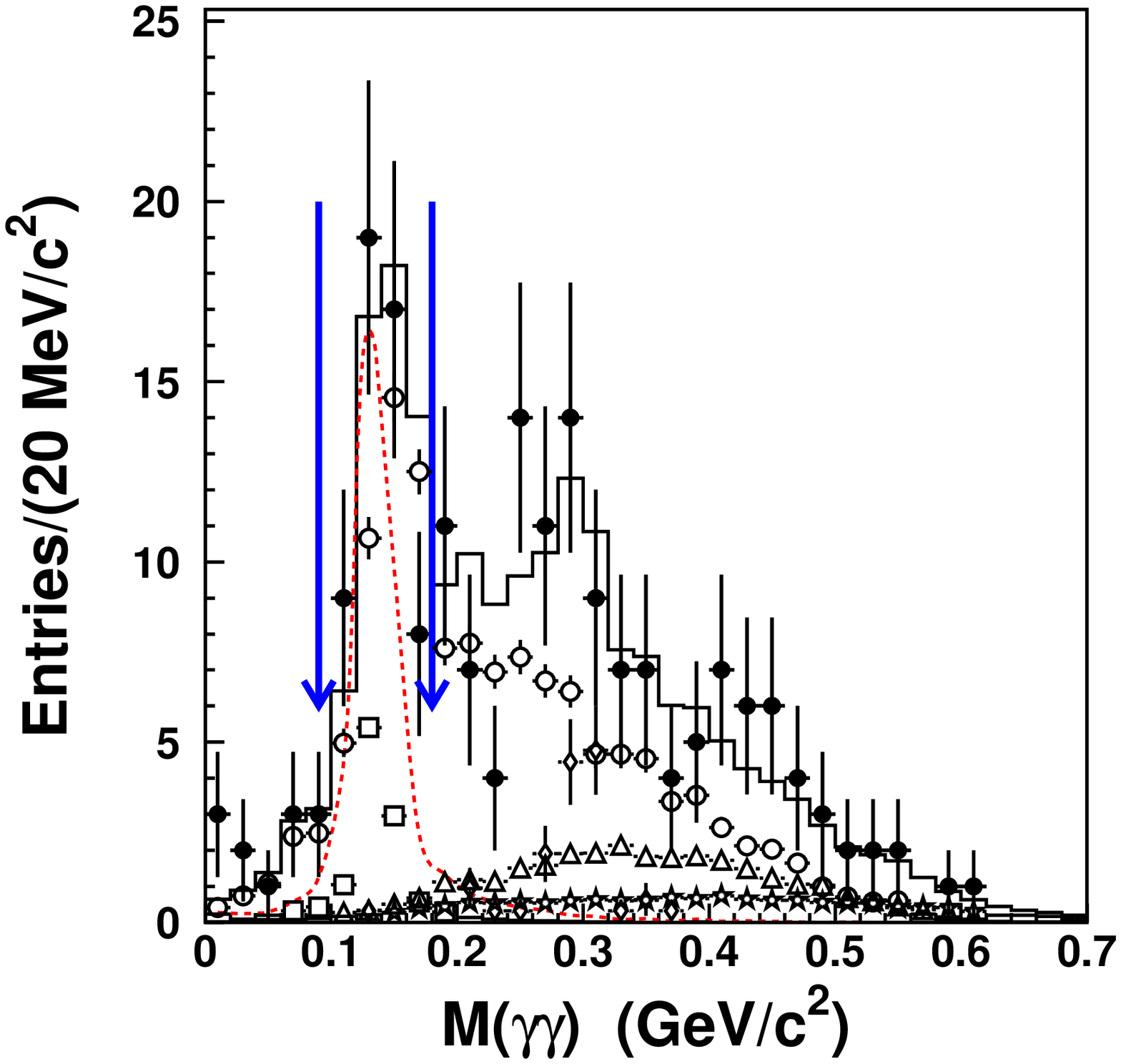,width=8cm,height=7cm}}
\caption{Invariant mass distribution of $M(\GG)$ for $\jpsi\ar\llb\GG$
  candidates (dots with error bars) and normalized backgrounds (solid
  histogram). The dashed curves shows the $\pi^0$ signal from MC
  simulated $\jpsi\ar\llp$. The arrows denote the region of the
  $\pi^0$ signal defined in the text. We use different histogram
  styles to indicate leading backgrounds from
  $\jpsi\ar\Sigma^0\pi^0\lmb~(+c.c.)$ (circles),
  $\jpsi\ar\Sigma^+\pi^-\lmb~(+c.c.)$ (squares),
  $\jpsi\ar\Xi^0\bar{\Xi}^0$ (triangles),
  $\jpsi\ar\Sigma(1385)^0\bar{\Sigma}(1385)^0$ (stars) and
  $\jpsi\ar\SSB$ (rhombi), which contribute $46.0 \pm5.4$,
  $11.2\pm1.3$, $1.9 \pm 0.4$, $1.0 \pm0.3$, and 0 events in the
  defined $\pi^0$ region.  }
\label{nnfig5}
\efg
\subsubsection{Background determination and upper limit on the number of signal events}
Using the branching fractions for $\jpsi\ar\Sigma^+\pi^-\lmb+c.c.$
measured above and branching fractions available in the
PDG~\cite{np12}, we obtain $29.2$, $14.3$, $14.2$, $125.0$, and
$11.9$ background events from $\jpsi\ar\Xi^0\bar{\Xi}^0$, $\jpsi\ar\SSB$,
$\jpsi\ar\Sigma(1385)^0\bar{\Sigma}(1385)^0$,
$\jpsi\ar\Sigma^0\pi^0\lmb~(+c.c.)$, and
$\jpsi\ar\Sigma^+\pi^-\lmb~(+c.c.)$ for the $\jpsi\ar\llp$ selection,
respectively. We also studied backgrounds from other
possible channels listed in the PDG~\cite{np12} that might contaminate
the $\pi^0$ signal, but their contamination was found to be negligible.
The histogram in Fig.~\ref{nnfig5} shows normalized backgrounds from
all background channels. The normalized $M(\GG)$
distribution of the background events is in reasonable agreement
with the data. The dashed line in the figure shows the $\pi^0$ signal
from MC simulated $\jpsi\ar\llp$. To estimate the expected number of
signal events, we define the $\pi^0$ mass region as
$|M(\GG)-M(\pi^0)|<0.045$ GeV/$c^2$, which is indicated in the figure
and selects most of the $\pi^0$ signal events. The numbers of $\pi^0$
events in the mass region are found to be $54.0 \pm 7.4$ and $60.1 \pm
9.5$ for data and normalized backgrounds, respectively. By using the
POLE method~\cite{np13,np14}, the upper limit on the number of $\pi^0$
events from $\jpsi\ar\llp$ is calculated to be 11.2 at the 90\%
confidence level (C.L.).
\bfg
\centerline{\psfig{file=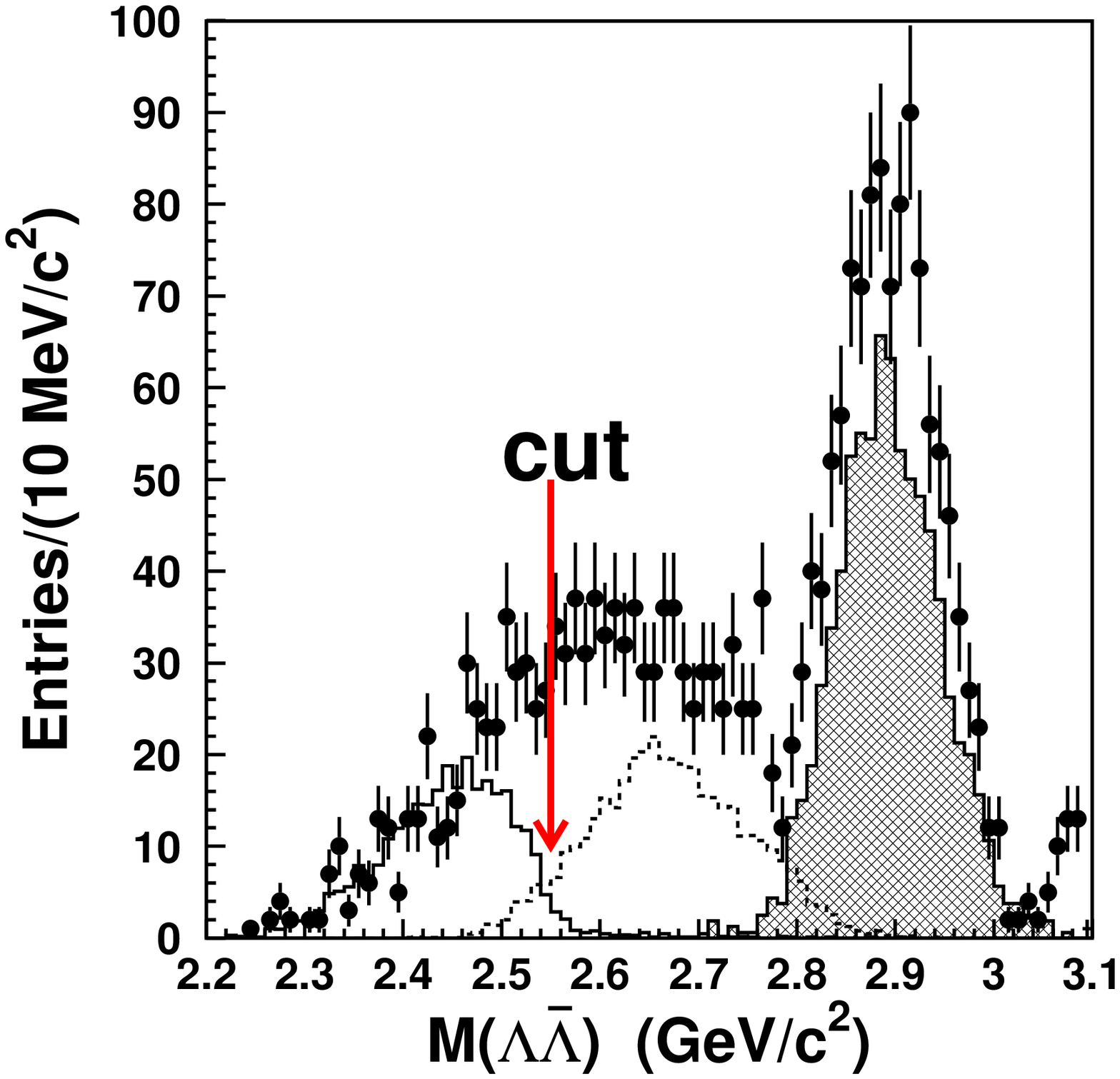,width=8cm,height=6cm}}
\caption{The $\llb$ invariant mass distribution for $\jpsi\ar\llb\GG$
candidates (dots with error bars), background from MC simulated
$\jpsi\ar\SSB$ (hatched histogram), background from MC simulated
$\jpsi\ar\Xi^0\bar{\Xi}^0$ (dashed histogram), and MC simulated
$\jpsi\ar\lle$ signal (solid histogram). The backgrounds are
normalized according to the branching fractions in the PDG and the $\lle$ signal is normalized using the branching fraction measured in this paper. The arrow indicates the $M(\llb)$ requirement, and events below the arrow are selected as $\jpsi\ar\lle$ candidates.}
\label{nnfig6}
\efg
\subsection{\boldmath $\jpsi\ar\lle$}
Candidate events with two or three good photons are selected, and the
$\chi^{2}_{4C}$ is required to be less than 15. Since the momenta of
$\lm$ and $\lmb$ are low in this channel, no requirement is made on
the decay lengths of $\lm$ and $\lmb$; otherwise the efficiency would
be extremely low. This is demonstrated in Fig.~\ref{nnfig5}, where a
decay length requirement is made and no $\eta$ signal is
seen. Figure~\ref{nnfig6} shows the $\llb$ invariant mass distribution
after the above selection. To remove the backgrounds from
$\jpsi\ar\SSB$ and $\jpsi\ar\Xi^0\bar{\Xi}^0$, events with $M(\llb)>
2.55$ GeV/$c^2$ are rejected, since for the signal process, they are
kinematically prohibited .  Dots with error bars in Fig.~\ref{nnfig9}
show the invariant mass of $M(\GG)$, and a clear $\eta$ signal is
observed.  \bfg
\centerline{\psfig{file=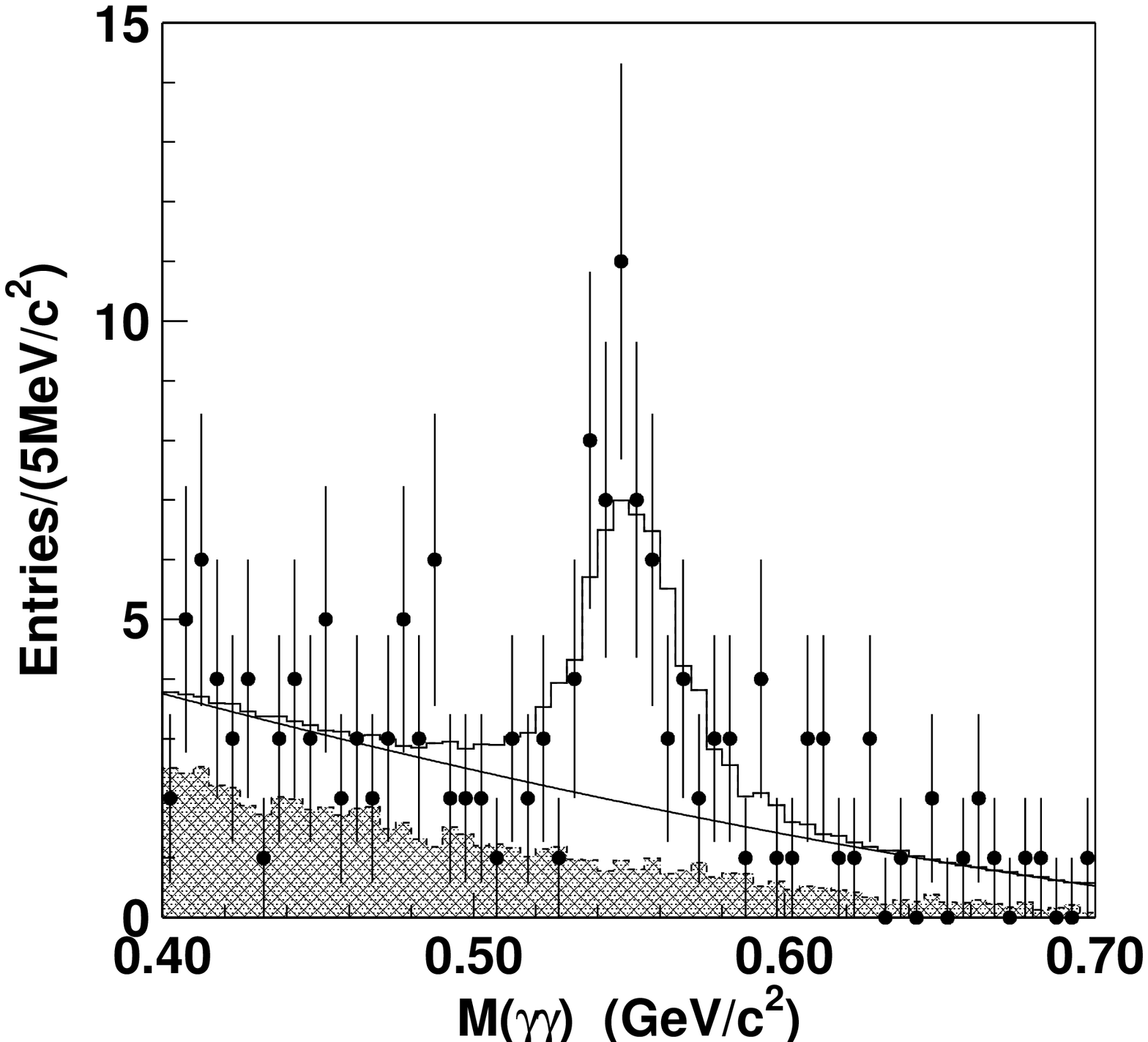,width=8cm,height=6cm}}
\caption{Fit to the $\GG$ invariant mass distribution of
$\jpsi\ar\llb\GG$ candidate events selected in Fig.~\ref{nnfig6}. Dots
with error bars are data, the hatched histogram is the normalized
background from all the channels considered, and the
solid histogram is the fit to data using a histogram of the signal
shape from MC simulation plus a second order polynomial for
background.}
\label{nnfig9}
\efg

To investigate possible backgrounds, we consider the following
channels with $\lm$ or $\Xi$ production: $\jpsi\ar\gamma\llb$, $\SSB$,
$\Sigma(1385)^0\bar{\Sigma}(1385)^0$, $\Xi^0\bar{\Xi}^0$,
$\Xi(1530)^0\bar{\Xi}^0$, $\Sigma^0\pi^0\lmb+c.c.$, and
$\Sigma^+\pi^-\lmb+c.c.$. Using available branching fractions of these decay modes,
we obtain 7.8, 27.6, 6.2, and 20.4 background events from $\jpsi\ar\Xi^0\bar{\Xi}^0$,
$\jpsi\ar\Sigma(1385)^0\bar{\Sigma}(1385)^0$, $\jpsi\ar\Xi(1530)^0\bar{\Xi}^0$, and $\Sigma^0\pi^0\lmb+c.c.$ in
the mass region $M(\GG)>0.4$ GeV/$c^2$, respectively. The background
contribution from the $\llb$ sidebands ($|M(\pi p)-1.141|<0.01$
GeV/$c^2$) is evaluated to be $3 \pm 2$ events. Contamination from
other possible channels listed in the PDG~\cite{np12} that might
contaminate the $\eta$ signal is negligible. The shaded histogram in
Fig.~\ref{nnfig9} shows the normalized backgrounds from the above
channels. We fit the $\GG$ invariant mass distribution with a MC
simulated signal shape and a second order polynomial background. The
fit yields $44 \pm 10$ events with a statistical significance of
4.8$\sigma$.

\subsection{\boldmath $\psp\ar\llp$ and $\lle$}
The selection criteria for these two decays are similar to those for
$\jpsi$ decays. A 4C kinematic fit to the hypothesis $\psp\ar\GG\pppr$
for candidate events with two good photons is performed, and the
$\chi^{2}_{4C}$ is required to be less than 15. Backgrounds from
$\psp\ar\pi^+\pi^-\jpsi$ are rejected with the requirement
$|M^{recoil}_{\pi^+ \pi^-}-M(\jpsi)|>0.04$ GeV/$c^2$, where
$M^{recoil}_{\pi^+ \pi^-}$ is the recoiling mass of $\pi^+
\pi^-$. Figure~\ref{nnfig12} (d) depicts the invariant mass distribution
of the charged tracks. The peak around 3.1 GeV/$c^2$ is from
$\psp\ar neutral+\jpsi$. In order to veto such background,
$|M(\llb)-M(\jpsi)|>0.05$ GeV/$c^2$ is required. Furthermore, to
suppress the background from $\psp\ar\SSB$ shown in
Fig.~\ref{nnfig12}, the invariant mass of $\llb$ is required
to be less than 3.3 GeV/$c^2$.
\bfg
\centerline{\psfig{file=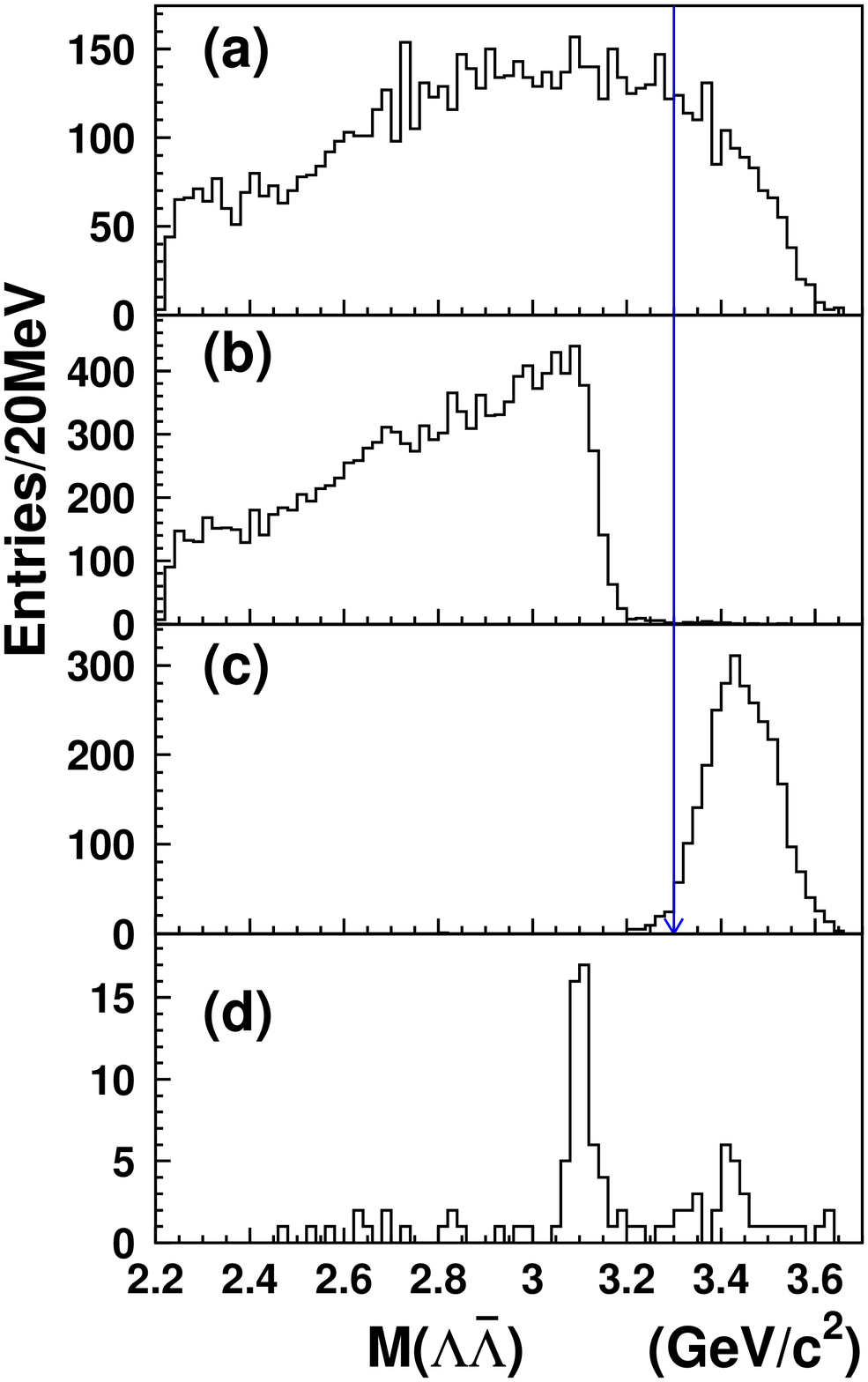,width=8cm,height=6cm}}
\caption{$M(\llb)$ distribution for MC simulated events: (a)
$\psp\ar\llp$, (b) $\psp\ar\lle$, and (c) $\psp\ar\SSB$. (d) The $M(\llb)$
distribution for data. The arrow denotes the
selection $M(\llb)<$ 3.3 GeV/$c^2$.}
\label{nnfig12}
\efg
\bfg
\centerline{\psfig{file=newfig1/nfig1010.epsi,width=8cm,height=6cm}}
\caption{The $\GG$ invariant mass distribution for candidate
$\psp\ar\GG\llb$ events. Dots with error bars are data, and the
histograms are MC simulated signal events. The arrows indicate the
signal region of $\pi^0$ and $\eta$ described in the text.}
\label{nnfig13}
\efg

Figure~\ref{nnfig13} shows the $\GG$ invariant mass distribution
after the above selection, and we see no significant
$\pi^0$ or $\eta$ signals. In order to estimate the number of signal events,
we define the signal regions as 0.09 $<M(\GG)<$ 0.18 and
0.50 $<M(\GG)<$ 0.60 (GeV/$c^2$) for $\pi^0$ and $\eta$,
respectively. The number of signal events is found to be 4 in both
regions. To estimate the backgrounds from the sidebands of
$\pi^0$ and $\eta$, (0.03 - 0.08) and (0.19 - 0.25) (GeV/$c^2$) are taken
as the sidebands of $\pi^0$, and (0.43 - 0.49), (0.61 - 0.67) (GeV/$c^2$)
are taken as the sidebands of $\eta$. The numbers of background events from
the sidebands of $\pi^0$ and $\eta$ are estimated to be 1 and
1.5. With the POLE method~\cite{np13,np14}, the upper limits on the
numbers of signal events at the 90\% C.L. are calculated to be 7.0 and
7.6, respectively.
\begin{table*}
\caption{Summary of systematic errors (\%).}
\bcl
\begin{tabular}{l|cc|cc|cc}\hline\hline
Source&$\jpsi\ar\llp$&$\jpsi\ar\lle$&$\psp\ar\llp$&$\psp\ar\lle$&$\jpsi\ar\splb$&$\jpsi\ar\sbpl$ \\ \hline
Tracking and PID&7.0&14.0&6.0&12.0&7.0&6.0\\
Photon efficiency&4.0&4.0&4.0&4.0&4.0&4.0\\
Kinematic fit&5.0&5.0&5.0&5.0&5.0&5.0\\
$\lm$ vertex requirement&3.7&-&-&-&-&-\\
Background shape&-&3.0&-&-&2.2&1.5 \\
Number of good photons&3.0&3.0&3.0&3.0&-&- \\
Total number of events&4.7&4.7&4.0&4.0&4.7&4.7\\
Total&11.6&16.6&10.1&14.5&10.8&10.1 \\ \hline
\end{tabular}
\label{syserrors}
\ecl
\end{table*}

\section{Systematic errors}
The systematic errors on the branching fractions are mainly from the
efficiency differences between data and MC simulation in the MDC
tracking, particle identification (PID), photon detection, kinematic
fitting, the $\lm$ vertex finding, and the decay length requirement
and the uncertainties on the
total number of $\jpsi$ and $\psp$ events.

The MDC tracking and particle identification (PID) systematic errors
are estimated from the difference of the selection efficiencies of
protons and antiprotons between data and MC simulation~\cite{np15}. The
efficiencies are measured using samples of $\jpsi\ar\pppr$ and
$\psp\ar\pppr$, which are selected using PID for three tracks,
allowing one proton or antiproton at a time to be missing in the
fit~\cite{np15}. The efficiency difference between data and MC
simulation for one proton is from 2\% to 5\% depending on the
proton momentum of the decay channels.

The photon detection efficiency is studied using $\jpsi\ar\rho^0\pi^0$
in Ref.~\cite{np16}. The results indicate that the systematic error is
about 2\% for each photon. Therefore, 4\% is taken as the systematic
error on the photon efficiency for all the decays.

\begin{table*}
\caption{Measured branching fractions or upper limits at 90\% confidence level (C.L.) for all the studied channels. Here, ${\cal B}(\Lambda\ar\pi^- p)=63.9\%$,
 ${\cal B}(\Sigma^+\ar\pi^0 p)=51.6\%$ and ${\cal
 B}(\eta\ar\GG)=39.4\%$ are taken from the PDG.}
\bcl
\begin{tabular}{l|c|c|c}\hline\hline
Channels&Number of events&MC efficiency(\%)&Branching fraction ($\times 10^{-4}$)\\ \hline
$\jpsi\ar\llp$&$<11.2$&0.75&$<0.64$\\
$\jpsi\ar\lle$&$44\pm 10$&$1.8$&$2.62\pm 0.60\pm 0.44$\\ \hline
$\psp\ar\llp$&$<7.0$&2.5&$<0.49$\\
$\psp\ar\lle$&$<7.6$&2.9&$<1.2$\\ \hline
$\jpsi\ar\splb$&$335 \pm 22$&2.3&$7.70\pm 0.51\pm 0.83$ \\
$\jpsi\ar\sbpl$&$254 \pm 19$&1.8&$7.47\pm 0.56\pm 0.76$ \\ \hline
$\jpsi\ar\splb+c.c.$& & &$15.17\pm 0.76\pm 1.59$ \\  \hline
\end{tabular}
\label{branresult}
\ecl
\end{table*}
The uncertainty due to the kinematic fit is studied using many
channels which can be selected purely without a kinematic
fit~\cite{np15,np16,np17}. It is found that the MC simulates the
kinematic fit efficiency at the 5\% level for almost all channels
tested. Therefore, we take 5\% as the systematic error due to the
kinematic fit.

The $\lm$ reconstruction systematic errors are
studied using $\jpsi\ar\llb$~\cite{np7,np8}. The $\lm$
secondary vertex finding gives a systematic error of 0.7\%
for each $\lm$ vertex, and the decay length requirement
contributes 1.7\%. The total percentage error arising from $\lm$ and
$\lmb$ vertex requirements is 3.7\%.

The systematic error of the background shape can be determined by
fitting the observed $\Sigma^+$, $\bar{\Sigma^-}$ and $\eta$ signal
events with different background shapes. For $\jpsi\ar\splb$
and $\jpsi\ar\sbpl$, the background shape in fitting the $\Sigma^+$
and $\bar{\Sigma^-}$ is changed to a second order polynomial. The
differences in the numbers of fitted $\Sigma^+$ and $\bar{\Sigma^-}$
events are found to be 2.2\% and 1.5\%, respectively. For
$\jpsi\ar\lle$, the background shape is changed from a second order
polynomial to a first order one, and the difference in the number of
fitted signal events is about 3\%.

The uncertainty caused by the requirement of two good photons is
estimated by considering the percentage of events without fake photons
in the sample of $\jpsi\ar\llb$. It is found that the difference in
the percentages of events without fake photons between data and MC
simulation is 3\%, which is taken as the systematic error for the
requirement of two good photons.

Finally, the results reported here are based on a total of 58 M
$\jpsi$ events and 14 M $\psp$ events. The uncertainties on the number
of $\jpsi$ and $\psp$ events are 4.7\% and 4.0\%, respectively. Table
~\ref{syserrors} lists the systematic errors from all sources. Adding
all errors in quadrature, the total percentage errors
range from 10\% to 17\% for all the studied decay channels.

\section{Results and Discussion}
Table ~\ref{branresult} lists the results for $\jpsi$ and $\psp$ decay
into $\llp$ and $\lle$, as well as $\jpsi\ar\splb+c.c.$. We also list the total branching fraction
for the conjugate modes, where the common systematic errors have been taken out. Except for
$\jpsi\ar\llp$ and $\jpsi\ar\splb+c.c.$, the results are first
measurements. Interestingly, the result of $\jpsi\ar\llp$
presented here is much smaller than those of DM2 and
BESI~\cite{np3,np4}. In previous experiments, the large contaminations
from $\jpsi\ar\Sigma^0\pi^0\lmb+c.c.$ and
$\jpsi\ar\Sigma^+\pi^-\lmb+c.c.$ were not considered, resulting in a
large value of branching fraction for $\jpsi\ar\llp$. The small
branching fraction of $\jpsi\ar\llp$ and relatively large branching
fraction of $\jpsi\ar\lle$ measured here indicate that the isospin
violating decay in $\jpsi$ decays is suppressed while isospin
conserving decays are favored, which is consistent with
expectation.

\section{Acknowledgment}
The BES collaboration thanks the staff of BEPC and computing center
for their hard efforts. This work is supported in part by the National
Natural Science Foundation of China under contracts Nos. 10491300,
10225524, 10225525, 10425523,10625524, 10521003, the Chinese Academy
of Sciences under contract No. KJ 95T-03, the 100 Talents Program of
CAS under Contract Nos. U-11, U-24, U-25, and the Knowledge Innovation
Project of CAS under Contract Nos. U-602, U-34 (IHEP), U-612(IHEP),
the National Natural Science Foundation of China under Contract
Nos. 10225522, 10491305 (Tsinghua University), MOE of China under
contract No. IRT0624 (CCNU), and the Department of Energy under
Contract No. DE-FG02-04ER41291 (U. Hawaii).

\end{document}